# Discovering Student Behavior Patterns from Event Logs: Preliminary Results on A Novel Probabilistic Latent Variable Model

Chen Qiao
Division of Information and Technology Studies
University of Hong Kong
Hong Kong
cqiao@hku.hk

Xiao Hu
Division of Information and Technology Studies
University of Hong Kong
Hong Kong
xiaoxhu@hku.hk

*Abstract*— Digital platforms enable the observation of learning behaviors through fine-grained log traces, offering more detailed clues for analysis. In addition to previous descriptive and predictive log analysis, this study aims to simultaneously model learner activities, event time spans, and interaction levels using the proposed Hidden Behavior Traits Model (HBTM). We evaluated model performance and explored their capability of clustering learners on a public dataset, and tried to interpret the machine recognized latent behavior patterns. Quantitative and qualitative results demonstrated the promising value of HBTM. Results of this study can contribute to the literature of online learner modelling and learning service planning.

## I. INTRODUCTION

Thanks to advances in information technology, automatically recording learner behaviors and tracing learner interactions with online materials have become unprecedentedly convenient and efficient. With such rich information, it is possible to discover the underlying behavior patterns and recognize the variations across them. Previous studies have obtained interesting findings by making sense of learner logs from different perspectives [1-6], among which one line of research is to induce [2] or model [1] abstracted behavior patterns behind fine-grained log traces. The necessity of a tractable number of discriminant hidden patterns lies in that they not only relieve the burden of usually intractable enumeration and interpretation of all the learner event trajectories [1], but also capture the manifold or most of the variations of the data with a limited set of variables. These benefits make them an effectiveness- and efficiency-balanced resource for researchers studying online education and platform vendors promoting educational services.

This study aims to simultaneously model learner events, time spans, as well as learner-resource interactions discretized at levels of mouse clicks and keyboard strokes in a unified probabilistic framework, which is derived from the Latent Dirichlet Allocation model (LDA) [7]. Originally invented for modelling documents, LDA can be trained to represent a document with a set of probabilistically weighted latent topics which are assumed to impact the writing of each observable word in this document. LDA has been successfully applied in text analysis [8-10], and derived to process other types of data such as images [11, 12] and gene sequences [13]. The same mechanism could also be leveraged to model the generation process of log events regarding a fixed number of hidden behavior patterns shared by all log traces (a sequence of log events created by a subject). Specifically, under the LDA model, the variations of event traces can be attributed to the different distributions of the underlying hidden behavior patterns.

The extension of LDA to model event logs in this study makes it possible to represent learner log traces with a finite set of hidden behavior patterns, and make these hidden patterns interpretable with the distributions of log events, timespans and interaction intensity (See section IV. A). Meanwhile, learners, being the owners of the event logs, can also be represented via the hidden behavior patterns, with similar representations indicating similarly behaved learners. We name the adapted model Hidden Behavior Traits Model (HBTM), as it is more intuitive for understanding its purpose. HBTM reduces the dimensionality of observed data to the finite number of hidden patterns (referred to as "traits" in this study), and this study will demonstrate how this approach can help analyze large quantity of log data.

The major contributions of this paper are twofold:

1) we propose HBTM which collectively models learner log events, event time spans and interaction levels in a unified probabilistic framework, which can serve as a basic analytic approach for future studies;

2) we implement, explore and interpret the HBTM in a public dataset, demonstrating its values in practice.

## II. RELATED WORKS

Student logs in the virtual learning environment record learner interactions with digital materials, from which learner behaviors and learning status can be inferred with special computational methods. To map the overwhelming volume of log data into a restricted space where interpretation and analysis are tractable, several studies have been devoted to discovering a constrained set of underlying hidden patterns that capture the greatest variation of log traces. One cluster of the methods assume specific stochastic processes for the variables, with certain probabilistic models. For example, [14] assumed the relationship between learner engagement and activities to follow a hinge-loss Markov random field model, and specified engagement to be a latent variable that impacts observable activities. Their modelling proved to be helpful for predicting student performance. Besides, [15] assumed the existence of subpopulations among students with different engagement levels and that they can be discovered with mixture of Gaussians (or equivalently, k-means clustering). Another cluster of methods, on the contrary, induce patterns by abstraction and generalization from observations without a formal definition of the processes for the variables. A representative method is the process mining [6] method, with which [2] abstracted out learning processes from students' accumulated session logs and calculated metric indices of process complexity. Their work offers a different angle of event log exploration and provides the community with helpful indices associated with course session difficultness and course achievements.

LDA has become a general modelling framework and been widely applied in various fields. Leveraging LDA in modelling human characteristics and behaviors has seen many innovative studies. For example, [17] treated character types in films as latent variables which are characterized by distinctive distributions over three lexical clusters, under which the film persona Zombie can be inferred from its distributions biased towards the agent terms of the words in the eat and kill classes, the patient terms of the words in the kill classes, and the object terms of the words in the dead class, echoing the fact that Zombies are usually described as being already dead, used to kill and eat living creatures and are likely to be killed by heroes. In a different setting, [18] adapted LDA to cope with clinical activity logs and timestamps of patients, and treated the latent variables as clinical pathway patterns. Likewise, [1] used LDA to model MOOC users and took the latent variables as use cases that explain the variation of learner behaviors, the clues from the use cases are greatly indicative of course certification.

This study assumes that the essential patterns of online learners can be captured by LDA-like probabilistic models, and will extend LDA to collectively fit event, time spans and resource interaction level in a unified model. This study is inspired by [18] and can be taken as a step further of [1].

### III. Hidden Behavior Traits Model

The plate notation for HBTM is illustrated in Figure 1. In contrast to LDA, the generative procedure of HBTM incorporates more steps, which is described as follows.

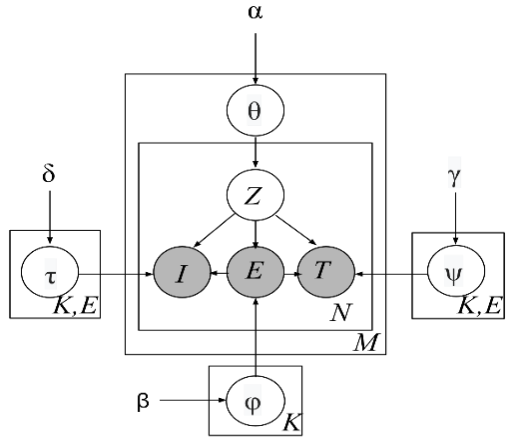

Figure 1: Plate notation for HBTM

In a log dataset of $M$ event traces where $K$ hidden behavior traits are assumed: first,

- parameters of $M$ Categorical distributions (i.e., Multinomial distributions with $n = 1$) $Cat(\theta_1),\ldots,Cat(\theta_M)$ over $K$ hidden traits are sampled from the Dirichlet prior distribution $Dir(\alpha)$;
- parameters of $K$ Categorical distributions $Cat(\phi_1),\ldots,Cat(\phi_K)$ over $E$ hidden traits are sampled from the Dirichlet prior distribution $Dir(\beta)$;
- parameters of $K * E$ Categorical distributions $\{Cat(\psi_{k,e}) | k = 1,\ldots,K, e = 1,\ldots,E\}$ over discretized time spans are sampled from the Dirichlet prior distribution $Dir(\psi)$;
- parameters of $K * E$ Categorical distributions $\{Cat(\tau_{k,e}) | k = 1,\ldots,K, e = 1,\ldots,E\}$ over mouse and keyboard interaction levels are sampled from the Dirichlet prior distribution $Dir(\tau)$;

To generate the $n$-th event log of the $m$-th trace, which is a tuple of event type, time span intervals and interaction intensity levels $(e_{m,n}, t_{m,n}, i_{m,n})$ (see section 3.1) instead of a single word in LDA, then,

- the Categorical distribution $Cat(\theta_m)$ first samples a hidden trait $z_{m,n}$,
- the corresponding Categorical distribution $Cat(\phi_{z_{m,n}})$ given $z_{m,n}$ is then selected to sample an event $e_{m,n}$,
- given event $e_{m,n}$, the corresponding $Cat(\psi_{z_{m,n},e_{m,n}})$ and $Cat(\tau_{z_{m,n},e_{m,n}})$ are selected to sample the time span $t_{m,n}$ and interaction level $i_{m,n}$ in parallel.

The sampling procedure repeats until all learners' log events are observed. The joint likelihood of the log data set $p(\phi_{1:K}, \theta_{1:M}, \psi_{1:K,1:E}, \tau_{1:K,1:E}, z_{1:M,1:N}, e_{1:M,1:N}, i_{1:M,1:N}, t_{1:M,1:N})$ is given in the following equation:

$$\prod_{i=1}^{K} p(\phi_i) \prod_{j=1,k=1}^{K,E} p(\psi_{j,k}) p(\tau_{j,k}) \{ \prod_{m=1}^{M} p(\theta_m)$$
$$\times \prod_{n=1}^{N} \{ p(z_{m,n}|\theta_m) p(e_{m,n}|\phi_{1:K}, z_{m,n})$$
$$\times p(t_{m,n}|\psi_{1:K,1:E}, e_{m,n}, z_{m,n}) p(i_{m,n}|\tau_{1:K,1:E}, e_{m,n}, z_{m,n}) \} \}$$

### IV. Experiment and Result

#### A. Dataset

We fit and evaluate our model on a public data set [2]. This data set contains student logs from 6 laboratory sessions in a first-year undergraduate digital electronics course. Data were collected from 115 subjects. Each session lasts over two hours and events such as studying materials, editing text editors, and accessing a digital simulator were all recorded together with the time duration and mouse and keyboard interactions. In addition to event logs, student scores of session assessments (except for the 1st session) and final exam scores were also available in the dataset. The original dataset contains in total 230318 logged events.

To enhance the quality of the dataset, we filtered out those transient log events lasting less than 1 second, and those seemingly "frozen" events lasting over 20 minutes (including outlier events). We also discretized time durations to seven-degree bins: (0, 9], (9, 15], (15, 30], (30, 60], (60, 600], (600, 1200] and (1200, l4000] in seconds; and mouse and keyboard interactions to 5 levels: [0, 2), [2, 3), [3, 6), [6, 16) and [16, 4779) total counts of mouse clicks and keyboard strokes, to simplify model computation. Also, there are 15 event types after aggregating events that belong to the same resource type (see appendix).

We preprocessed all the log data such that each student log trace in each session can be analogized to a document in LDA, and the 3-tuples (event, time span degree, interaction intensity level) are the "words" in the "documents".

*B. Experiment Setup*

We implemented our model in Python with the probabilistic programming package pymc3 featuring its implementation of advanced Markov chain Monte Carlo (MCMC) samplers. We compiled the likelihood computation method of the data based on the equation given in section 2, and leveraged the package's Metropolis-Hastings sampler for parameter inference. Model fitting was conducted on each of the 6 sessions, with numbers of hidden traits being K=5, 10, 15 and 20 respectively. The experiment was conducted on a remote CentOS 6.8 server equipped with two E5-2630v4 2.20 GHz CPUs and 32GB memory, and the pymc3 package was run at the CPU-version backend.

*C. Results and Analysis*

After fitting the model, every student in a session can be characterized with different distributions over the latent behavior patterns/traits. To understand whether the representation can distinguish learners in terms of their learning performances, we conducted K-means clustering to group the students into 2 clusters based on the latent behavior patterns/traits only. Independent samples t-tests were then conducted to compare the two groups in terms of their session-wise assessment grades (SA, scored 0-5), session-aligned problem grades in the final exam (S.FE, scored with various numerical grades), and the total final exam grades (FE, 0-100). The results are displayed in Table I, where only the significant ($p < 0.05$) models (with the number of latent behavior traits K designated) are displayed.

It is found that several model options in sessions 3, 4 and 5 captures the variances in more than one kind of grades: the 20- and 5-HBTM in session 3 and 4 can significantly distinguish students in terms of both the immediate session exercise scores (SA) and the session-corresponding exam problem scores in the final exam (S.FE), while the 10-HBTM in session 5 learned the information to distinguish the session-aligned final exam problem scores (S.FE) and the total final exam score (FE). However, no models managed to cope with all types of grades.

To explore the relationship between individual latent behavior traits and student grades, we computed the Pearson's correlation coefficients between each hidden trait in each model on each session and student grades. Latent traits that were significantly correlated ($p < 0.5$) with the grades are listed in Table II, with their model and session combinations designated and the coefficient signs marked in brackets. Most models contained at least one latent trait that was either positively or negatively correlated to some types of student grade.

Moreover, of all the options in Table II, it is interesting to find that the 20-HBTM on session 6 (last row of Table II) is particularly well-tuned to have both negative (T4, T13) and positive (T19) hidden behavior traits that were correlated with all the three grade types. To take a closer look at what the traits are like, we visualized Trait 13 (negatively correlated with all three grade types) and Trait 19 (positively correlated with all the three grade types) in Figure 2.

Figures 2. (a) and (b) show the probability distributions over the 15 events given Trait 13 and 19 respectively. The probability mass is more evenly distributed in Figure 2. (a), in contrast to the distribution in Figure 2. (b), which is more biased to event 9. Besides, events 14 and 15 in Figure 2. (a) captures a much greater total mass than those in Figure 2. (b) (almost zero). A closer examination of the event description in the appendix helps us recognize that event 14 and 15 are more likely to be off-topic events, while event 9 implies that the student is taking initiative trials to set up and observe

TABLE I. MODEL OPTIONS SIGNIFICANTLY DISTINGUISHING LEARNERS IN GRADES OF THREE TYPES

|  | SA | S.FE | FE |
| --- | --- | --- | --- |
| Session 1 | - | - | K=20 |
| Session 2 | - | - | - |
| Session 3 | K=5,10,20 | K=20 | - |
| Session 4 | K=5 | K=5,10 | - |
| Session 5 | - | K=10 | K=10 |
| Session 6 | - | - | - |

TABLE II. LATENT TRAITS SIGNIFICANTLY CORRELATED WITH GRADES OF THREE TYPES (T# IS THE TRAIT ID)

| Session /Model | SA | S.FE | FE |
| --- | --- | --- | --- |
| Sess 1 K 15 | - | (-)T1 | - |
| Sess 2 K 5 | (+)T2 | - | - |
| Sess 2 K20 | - | (-)T1 | - |
| Sess 2 K 15 | (-)T13 | (-)T2 | (+)T5 |
| Sess 2 K10 | - | (+)T1 | (+)T4 |
| Sess 3 K 5 | (+)T3 (-)T4 | - | - |
| Sess 3 K 10 | (+)T1,T2 (-)T7 | (-)T8 | (+)T1 |
| Sess 3 K 15 | (+)T6 (-)T3,T4,T13 | (-)T11 | (-)T4,T13 |
| Sess 3 K 20 | (+)T1,T5,T19 (-)T4,T8 | (-)T3 | (+)T1 (-)T5 |
| Sess 4 K 5 | (+)T5 | - | - |
| Sess 4 K 15 | - | (+)T2 | - |
| Sess 4 K 20 | (+)T11 | - | (+)T11 |
| Sess 5 K 5 | (-)T5,T10 | - | - |
| Sess 5 K 10 | - | (-)T2 | - |
| Sess 5 K 20 | (+)T16 | (+)T15,T16 | - |
| Sess 6 K 15 | (+)T15 (-)T4 | (+)T12,T15 | (+)T15 |
| Sess 6 K 20 | (+)T19 (-)T4,T13 | (+)T19 (-)T4,T13 | (+)T19 (-)T4,T12,T13 |

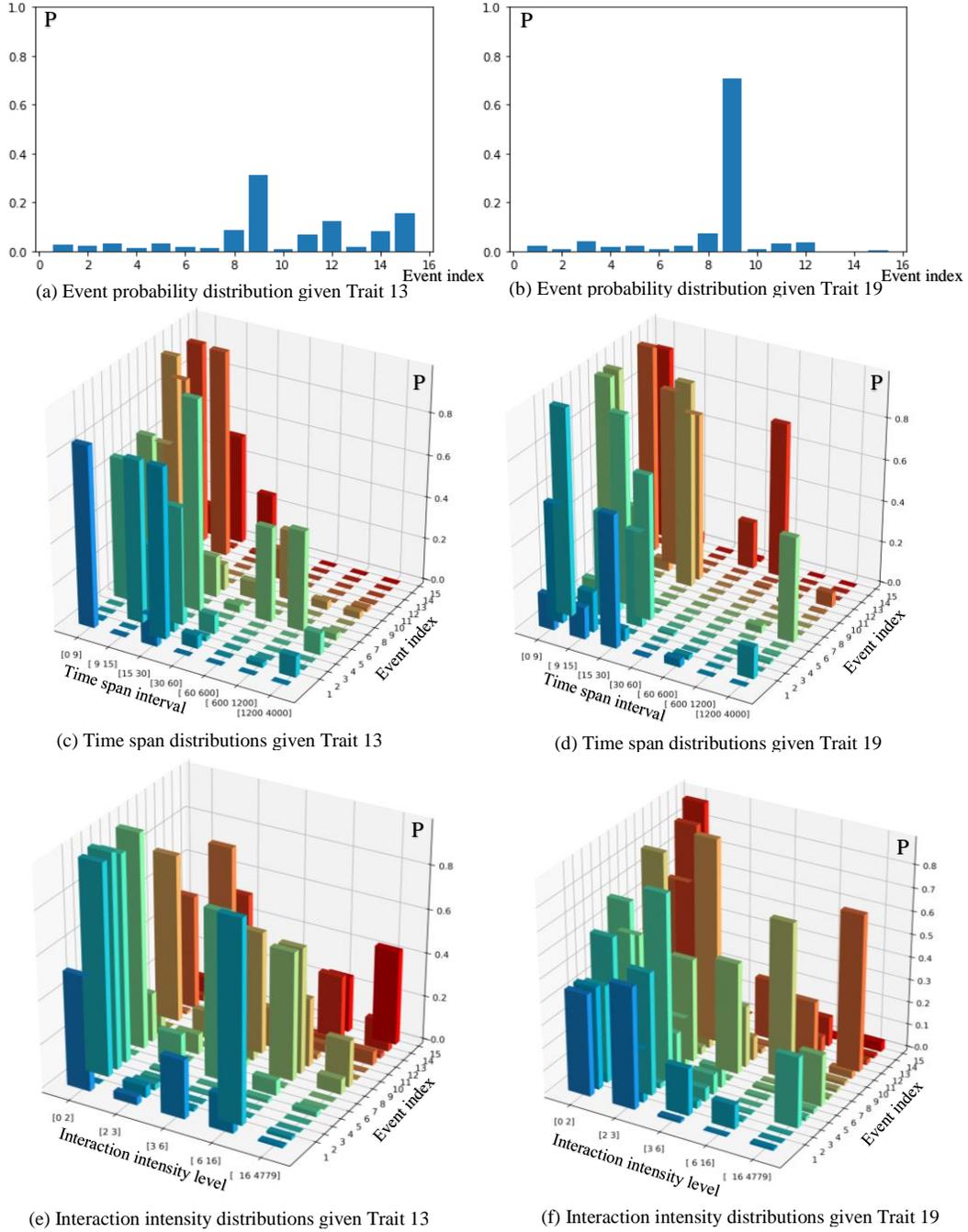

Figure 2. Event, time span & interaction intensity distributions given Hidden Trait 13 and 19 of the 20-HBTM on Session 6
(bars in one color corresponds to the probabilistic distribution given an event)

simulation experiments, though the setup event usually takes less time (most of the probability mass of event 9 are aggregated in the (0, 9] seconds bin as shown in Figure 2. (d). In addition, by comparing Figure 2. (c) and Figure 2. (d) we discover that, given event 15 where one visits other websites, he/she impacted by Trait 13 is more likely to spend more time (9~30 seconds with probability around 0.8) than those impacted by Trait 19 (0~9 seconds with probability close to 0.9). Likewise, much more mouse clicks and keystrokes are likely to be made in the off-topic events (Events 14 and 15) given Trait 13 than those given Trait 19 (see Figure 2. (e) and Figure 2. (f), the probability mass is over 0.6 in the [6,4779) interval for the former and almost 0 for the latter). From the visual clues, we can reasonably interpret Trait 13 as a less-focused behavior pattern, while Trait 19 is a more focused behavior pattern and may imply in-depth learning.

Nonetheless, a log trajectory is influenced by the combined influences of K hidden traits together for a specific subject. In other words, accompanied by other hidden behavior traits with different likelihoods, Trait 13 and Trait 19 are likely to play different roles in different learners' learning behaviors.

V. DISCUSSION AND FUTURE WORK

In this initial work, we proposed the Hidden Behavior Traits Model as a unified framework of modelling learner behaviors, event time spans and interaction intensity. We implemented and fitted this model on a public dataset. Starting as an unsupervised procedure, this model recognized patterns from learner logs. Quantitative results demonstrate that a proportion of these patterns capture information in distinguishing students in terms of assessment and exam scores, even though the unsupervised learning was not targeted at predicting grades. Inspection of two visualized hidden patterns revealed what had been learnt by the models, which helped us interpret logged behaviors of students. In summary, the HBTM is a promising model in terms of effectiveness and interpretability.

Although in this initial work we explored the hidden behavior patterns regarding learner grades, these patterns can explain other aspects of learning or learners apart from performances. Since the model is unsupervised, it is also possible to capture variance of the data in other aspects, such as demographic background and affective factors (e.g., motivation). Expanding the model's scope of application would be part of future work.

On the other hand, HBTM and the original LDA ignore event orders of the log trajectories, while it is possible that adjacent behaviors may have certain kinds of relations that are not captured by this type of models. Therefore, adapting or switching to other models which can capture local or global event orders can be another direction of future research.

APPENDIX: DESCRIPTION FOR LOG EVENTS

| Index | Description |
| --- | --- |
| 1 | It indicates that a student is studying / viewing the content of a specific exercise. |
| 2 | It indicates that the student is working on a specific exercise inside the Deeds simulator (Digital Circuit Simulator). |
| 3 | This shows when the student is on Deeds simulator but it is not clear what exercise he is working on. |
| 4 | It contains other activities related to Deeds, for instance when the students save circuit image or export VHDL. |
| 5 | when the student is writing the results of his work to submit later to the instructor. The students use a text editor (Word, Office, etc.) to answer to the questions and explain the solution they found through Deeds simulator. |
| 6 | It indicates that the student is working on an exercise in the text editor but it is not clear which exercise it is. |
| 7 | It shows that the student is using the text editor but not on exercises, this can contain other activities related to the text editor. |
| 8 | When the students use 'Simulation Timing Diagram' to test the timing simulation of the logic networks, while using the Deeds simulator. It also contains these components: "Input Test Sequence" and "Timing Diagram View Manager ToolBar". |
| 9 | Deeds simulator, Simulation Timing diagram, and FSM contain the properties window, which allows to set all the required parameters of the component under construction. For instance, the Properties can contain: "Switch Input", "Push-Button", "Clock properties", "Output properties", "textbox properties". All are labeled as 'Properties'. |
| 10 | The student is viewing some materials relevant to the course (provided by the instructor). |
| 11 | When the student is working on a specific exercise on 'Finite State Machine Simulator'. |
| 12 | When the student is handling the components of Finite State Machine Simulator. |
| 13 | Students are using Aulaweb as a learning management system (based on Moodle) which is used for the course of digital electronics at the University of Genoa. In Aulaweb, the students might access the exercises, download them, upload their work, check the forum news, etc. |
| 14 | When the title of a visited page is not recorded. |
| 15 | When the student is not viewing any pages described above, then the activity is assigned 'Other'. This includes, for majority of the cases, the student irrelevant activity to the course (e.g. if the student is on Facebook). |